\documentclass[journal]{IEEEtran}
\usepackage{cite}
\usepackage{times}
\usepackage{epsfig}
\usepackage{graphicx}
\usepackage{amsmath}
\usepackage{amssymb}
\usepackage{amsthm}
\usepackage{bm}
\usepackage{epstopdf}
\usepackage{url}
\usepackage{multirow}
\usepackage{booktabs}
\usepackage{tabularx}
\usepackage{threeparttable}
\usepackage[noend]{algpseudocode}
\usepackage{algorithmicx,algorithm}
\usepackage[caption=false,font=footnotesize]{subfig}
\newcommand{\visplaceholder}[2]{%
  \fbox{\parbox[c][#1][c]{0.95\linewidth}{\centering\footnotesize #2}}%
}
\newcommand{\optionalvis}[3][]{%
  \IfFileExists{#2}{\includegraphics[#1]{#2}}{\visplaceholder{#3}{Replace with visualization file\\\texttt{#2}}}%
}

\newcommand{\E}{\mathbb{E}}
\newcommand{\Tr}{\operatorname{Tr}}

\newcommand{\cov}{\operatorname{Cov}}
\newcommand{\f}{\mathbf{f}}
\newcommand{\uv}{\mathbf{u}}
\newcommand{\M}{\mathbf{M}}
\newcommand{\W}{\mathbf{W}}
\newcommand{\Sigm}{\boldsymbol{\Sigma}}

\begin{document}

\title{ChWDTA: Channel-wise Wavelet-Domain Transformer Attention and Entropy Modeling for Learned Image Compression}

\author{Haisheng~Fu,
        Runyu~Yang,
        Feng~Ding,
        Siyu~Zhu,
        Jie~Liang$^*$,
        Xiaoxiao~Li,
        Zhenman~Fang,
        and~Jingning~Han%
\thanks{Haisheng~Fu and Xiaoxiao~Li are with the Electrical and Computer Engineering Department, The University of British Columbia, Vancouver, BC, Canada 
(e-mails: haisheng.fu@ubc.ca; xiaoxiao.li@ece.ubc.ca).}%
\thanks{Runyu~Yang, Feng~Ding, and Zhenman~Fang are with the School of Engineering Science, Simon Fraser University, Burnaby, BC, Canada 
(e-mails: runyuy@sfu.ca; feng\_ding@sfu.ca; zhenman@sfu.ca).}%
\thanks{Jie~Liang and Siyu~Zhu are with the School of Electronic Science and Technology, Eastern Institute of Technology, Ningbo, China (e-mail: jliang@eitech.edu.cn; syzhu@eitech.edu.cn). $^*$Corresponding author: Jie~Liang.}
\thanks{Jingning~Han is with Google LLC, Mountain View, CA, USA 
(e-mail: jingning@google.com).}%
\thanks{This work was supported by the Natural Sciences and Engineering Research Council of Canada (RGPIN-2020-04525), Google Chrome University Research Program, NSERC Discovery Grant RGPIN-2019-04613, DGECR-2019-00120, Alliance Grant ALLRP-552042-2020; CFI John R. Evans Leaders Fund; MITACS Elevate Postdoc grant.}%
}

\markboth{JOURNAL OF LATEX CLASS FILES, VOL. 14, NO. 8, AUGUST 2021}%
{Fu \MakeLowercase{\textit{et al.}}: ChWDTA: Channel-wise Wavelet-Domain Transformer Attention and Entropy Modeling for Learned Image Compression}

\maketitle

\begin{abstract}
State-of-the-art learned image compression (LIC) schemes are increasingly based on hybrid CNN--transformer architectures. To further improve rate--distortion performance, we introduce channel-wise wavelet transforms into both the transformer and entropy-coding components. First, we propose a channel-wise wavelet-domain transformer attention (ChWDTA) mechanism. ChWDTA keeps the efficient windowed spatial self-attention used in modern LIC backbones, but computes the Q/K/V projections on channel-wise wavelet-transformed features before mapping the attention output back with the inverse transform. The resulting Channel-wise Wavelet-Domain Transformer Block (ChWDTB) therefore preserves the spatial tokenization pattern of windowed attention while sparsifying the channel covariance seen by the attention projections. Second, in the entropy-coding stage, we introduce a channel-wise wavelet packet (ChWP) decomposition that produces four equal-sized subbands, which better fit channel-wise slice-based autoregressive entropy modeling. When each channel-wise subband is divided into two slices, we use eight slices for entropy coding. With this configuration, the proposed scheme obtains BD-rate reductions of $-17.82\%$, $-19.15\%$, and $-22.56\%$ on the Kodak, CLIC Professional Validation, and Tecnick test sets, respectively. Even when each channel-wise subband is coded as a single slice, the scheme still retains most of the coding gains with lower complexity. The results confirm the advantage of introducing wavelet transform in the CNN-transformer-based LIC schemes. The source code is available at \url{https://github.com/fengyurenpingsheng/ChWDTA-Learned-image-compression}.
\end{abstract}

\begin{figure}[t]
  \centering
  \includegraphics[width=\linewidth]{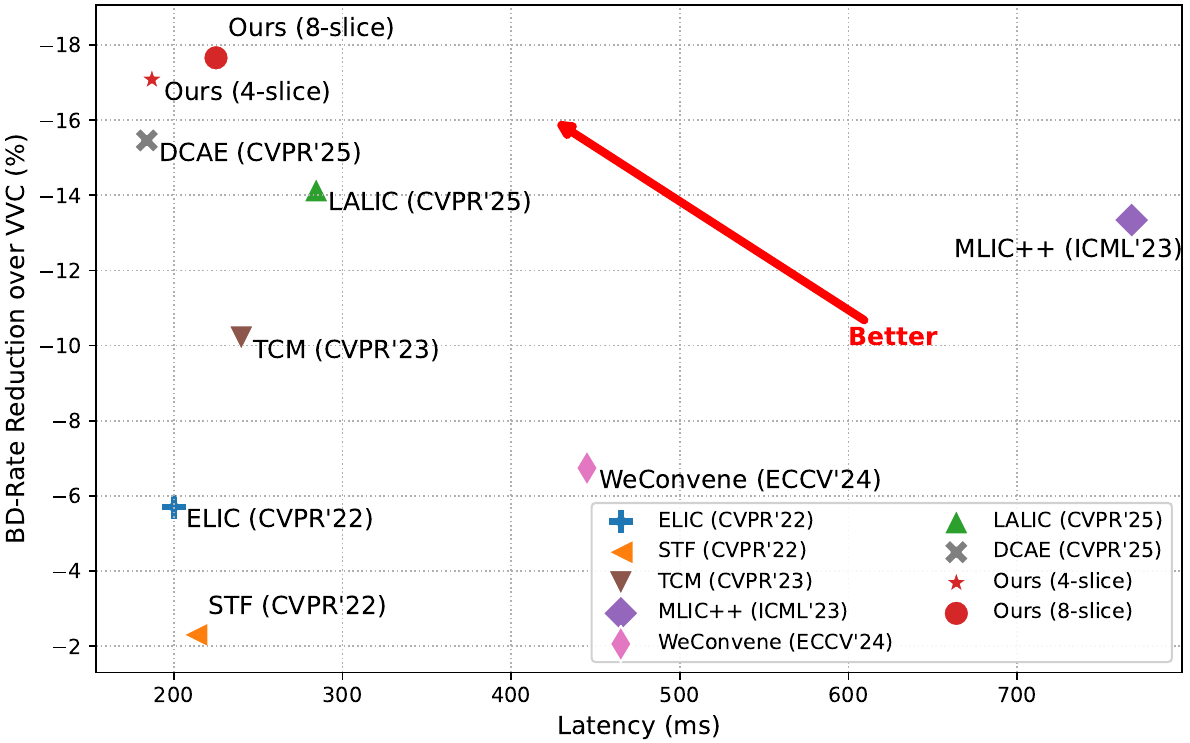}
  \caption{Rate--complexity performance of different methods on the Kodak test set. The x-axis is the total latency, including both encoding and decoding time. The y-axis reports BD-rate reduction relative to the VTM~9.1 VVC anchor. The labels ``Ours (4-slice)'' and ``Ours (8-slice)'' correspond to the proposed 4-slice and 8-slice schemes, respectively. The 4-slice model targets the low-latency regime with a favorable BD-rate--latency trade-off, while the 8-slice model further improves BD-rate with additional latency.}
  \label{fig:latency_bdrate}
\end{figure}

\section{Introduction}
\label{sec:intro}
 
Learned image compression (LIC) has rapidly advanced in recent years, surpassing the H.266/VVC standard in terms of both PSNR and MS-SSIM \cite{cheng2020,GLLMM,Liu_2023_CVPR,FAT,Feng2025LALIC,Lu_cvpr_2025,Li_2025_ICCV}.

Most LIC approaches adopt an end-to-end variational formulation in which an encoder--decoder network, a differentiable quantization proxy, and a learned entropy model are optimized jointly under a rate--distortion (R--D) objective. Since the introduction of the hyperprior~\cite{Variational}, subsequent work has progressively improved R--D performance through stronger backbones and increasingly expressive priors~\cite{Joint,jiang2023mlic,He_2022_CVPR,MambaVC,Lu_cvpr_2025, He_2021_CVPR, GLLMM, Entroformer_2022_ICLR}. In particular, hybrid CNN--transformer designs~\cite{Liu_2023_CVPR,Feng2025LALIC,Lu_cvpr_2025} have achieved state-of-the-art performance by combining the local modeling ability of convolutions with the contextual modeling capacity of self-attention.

Despite these advances, modern LIC methods still rely primarily on learned nonlinear transforms, while explicit transform structures such as DCT- or wavelet-like decompositions play a less central role than they do in classical transform coding. Classical codecs commonly apply fixed linear transforms, such as the DCT or wavelets, before quantization and entropy coding in order to reduce statistical dependence and concentrate signal energy. In many learned codecs, however, such reorganization is mainly left to the nonlinear analysis transform and is learned implicitly from data. As a result, the resulting latent representations may still contain substantial second-order dependence across channels. This residual channel dependence can limit both the transform backbone and the entropy model. For windowed spatial attention, the bilinear attention kernel has to operate in a channel coordinate system with diffuse cross-channel coupling. For slice-based entropy modeling, a diagonal conditional likelihood must approximate latent variables whose conditional covariance is not fully diagonal. These mismatches can increase the rate--distortion cost and motivate a structured channel-wise reparameterization before attention and entropy coding.

Wavelets have recently been reintroduced into LIC in several different ways. Some methods replace spatial sampling with learned wavelet-like transforms to build multiscale features \cite{Ma_2022_PAMI}. Others insert fixed or trainable wavelet transforms into the analysis or hyperprior pathway to promote frequency sparsity \cite{sahin2023learnedDWT,xue2022aiwave}. Another line of work performs convolution or entropy coding in the wavelet domain \cite{Fu_EECV}. These studies confirm that wavelet priors remain useful in LIC, but they do not fully explain how a channel-wise transform should interact with the transformer blocks used in current hybrid codecs.

Two issues are particularly relevant to our setting. First, most transformer modules in LIC are designed for spatial tokens. If a wavelet transform is inserted along the channel axis before the attention layer, it is important to explain what exactly changes: the resulting module is not a pure channel-attention transformer, but neither is the transform a meaningless preprocessing step. Second, existing wavelet-based entropy models often produce subband layouts that are not naturally aligned with slice-based channel-wise autoregressive modeling (ChARM)~\cite{channel,Liu_2023_CVPR,Lu_cvpr_2025}, making it difficult to exploit the transform structure inside modern lightweight entropy models.

To address these issues, we propose a channel-wise wavelet-domain transformer attention (ChWDTA) mechanism. ChWDTA denotes the attention operation whose Q/K/V projections are computed after a channel-wise wavelet transform. We further build a \emph{Channel-wise Wavelet-Domain Transformer Block (ChWDTB)} by wrapping ChWDTA with normalization, the channel-wise wavelet transform and its inverse, and a feed-forward module. In the entropy-coding stage, we introduce a two-level \emph{Channel-wise Wavelet Packet (ChWP)} decomposition that generates four equal-sized channel subbands, which are then split into balanced channel slices matched to slice-based coding.

The main contributions are summarized below.

\begin{itemize}

\item We wrap windowed spatial self-attention with a channel-wise wavelet transform and its inverse. The tokenization pattern remains spatial, but the Q/K/V projections are computed in a wavelet-transformed channel coordinate system. This makes the resulting ChWDTB a wavelet-domain spatial-attention block rather than a pure channel-attention layer.

\item We give a covariance-sparsification interpretation of the channel-wise wavelet transform as an invertible preconditioner of the attention kernel. The analysis clarifies that the gain should be understood under restricted modeling assumptions, such as finite-capacity attention and diagonal slice-wise entropy models. It also explains why the backbone attention path and the entropy path need not share the same wavelet parameters.

\item We introduce a two-level ChWP transform before entropy coding. Decomposing both smooth- and detail-branch channel groups in the second stage produces four equal-sized subbands, which are naturally aligned with slice-based entropy modeling and can reduce the residual covariance coupling left to the slice predictors.

\item We integrate ChWDTB into the main backbone, entropy path, and hyperprior, and evaluate two entropy layouts. The default 8-slice model obtains the most favorable BD-rate values among the compared methods in our evaluation: $-17.82\%$, $-19.15\%$, and $-22.56\%$ on Kodak, CLIC Professional Validation, and Tecnick, respectively. The 4-slice variant provides a lower-complexity operating point, but still achieves very competitive BD-rate reductions of $-17.24\%$, $-18.42\%$, and $-21.71\%$. These results show that the proposed method offers a favorable rate--complexity trade-off, as highlighted by the latency--BD-rate comparison in Fig.~\ref{fig:latency_bdrate}.

\end{itemize}

\begin{figure*}[t]
  \centering
  \includegraphics[scale=0.51]{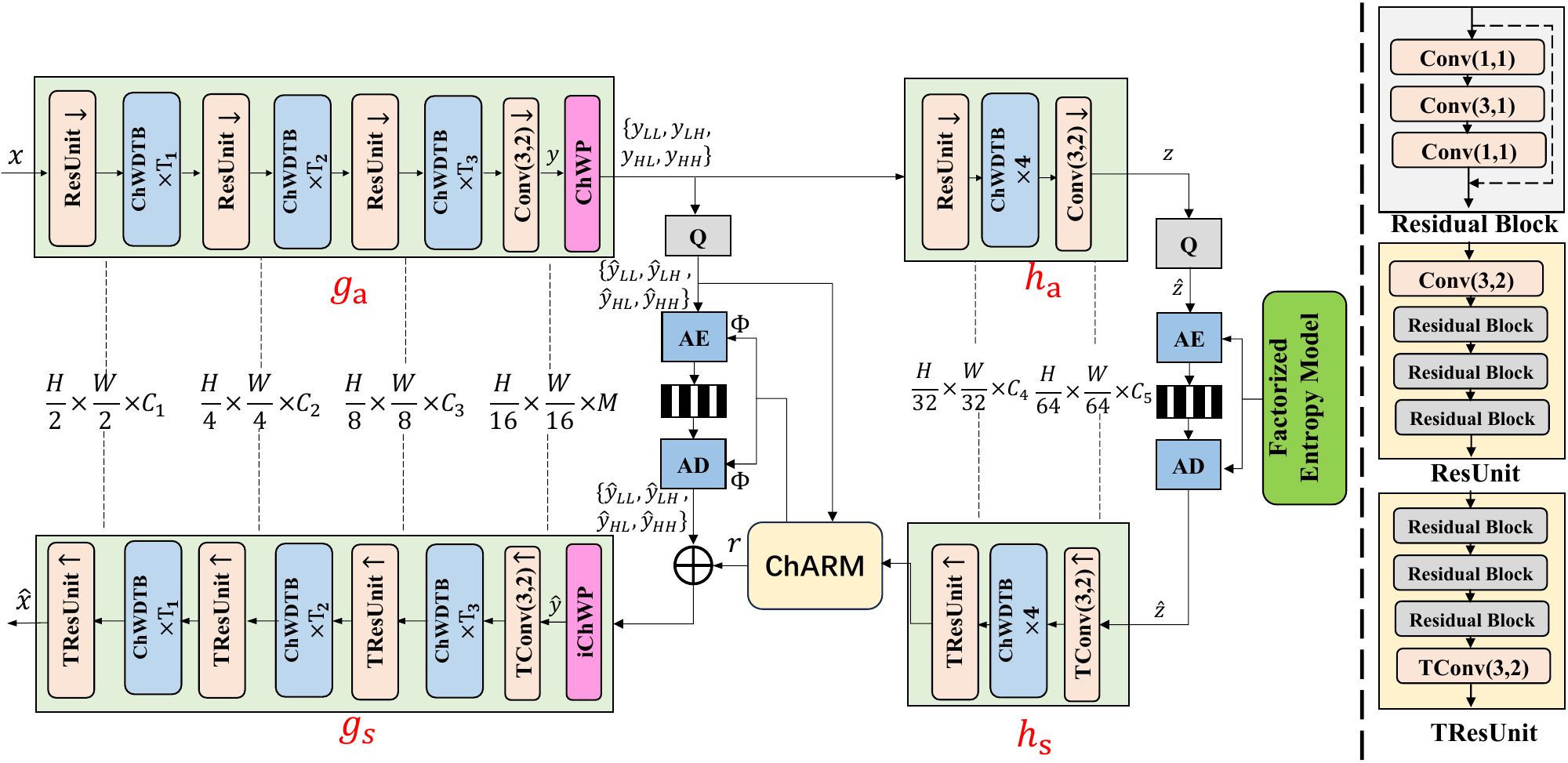}
  \caption{The overall architecture of the proposed scheme. The details of the ChWDTB and the ChARM module are shown in Fig.~\ref{fig:transformer} and Fig.~\ref{fig:charm_overall}, respectively. \texttt{Conv(3, s)} denotes a convolutional layer with a $3 \times 3$ kernel size and stride $s$, while \texttt{TConv(3, s)} denotes a transposed convolutional layer. Dashed shortcut connections indicate changes in tensor size. \texttt{AE}, \texttt{AD}, and Q represent the Arithmetic Encoder, Arithmetic Decoder, and quantization, respectively. The symbols \textbf{$T$} and \textbf{$C$} will be defined in Sec.~\ref{experiments}. }
  \label{fig:framework}
\end{figure*}

\section{Related Work}

\textbf{Learned image compression.} Learned image compression (LIC) optimizes nonlinear analysis/synthesis transforms and learned entropy models under a rate--distortion objective. Factorized priors and hyperpriors~\cite{end_to_end,Variational} established the variational coding framework by providing spatially adaptive likelihood estimates for latent representations. Later methods improved the likelihood model with autoregressive, hierarchical, mixture, and attention-based dependencies~\cite{Joint,cheng2020}. Subsequent works further strengthen this line from complementary directions, including Gaussian--Laplacian--logistic mixture likelihoods~\cite{GLLMM}, asymmetric efficient architectures~\cite{Asymmetric_Fu}, multi-resolution octave-based residual modeling~\cite{Mahammand_media}, nonlocal or attention-based entropy models~\cite{Li_Nonlocal_entropy}, and wavelet-like invertible transforms~\cite{Ma_PAMI}.

Recent high-performing LIC methods can be roughly divided into two groups. The first group improves the transform backbone. TCM~\cite{Liu_2023_CVPR} combines convolutional and transformer branches, FAT~\cite{FAT} introduces frequency-decomposition window attention, and LALIC~\cite{Feng2025LALIC} uses a Bi-RWKV-based linear-attention design for a favorable complexity--accuracy trade-off. State-space variants, including MambaVC~\cite{MambaVC} and CASSIC~\cite{Qin_2025_ICCV}, explore long-range modeling with reduced memory cost. The second group focuses on stronger entropy modeling. MLIC++~\cite{jiang2023mlicpp} uses multi-reference context modeling, DCAE~\cite{Lu_cvpr_2025} introduces a dictionary-based cross-attention entropy prior, HPCM~\cite{Li_2025_ICCV} progressively fuses hierarchical contexts, and MLICv2~\cite{jiang2025mlicv2} combines transform refinement, global context prediction, and instance adaptation. Practical learned codecs have also received increasing attention, including perceptual/runtime-aware design, region-wise codec selection, and image-specific overfitted coding~\cite{tatwawadi2026practicalLIC,blard2026spatialCompetition,ladune2026coolchic}. These works are complementary to ours. They mainly increase the capacity or efficiency of the backbone/context model, whereas our focus is a structured, perfect-reconstruction change of channel coordinates that is shared by attention and entropy coding.

\textbf{Frequency- and wavelet-aware learned compression.} Classical image codecs explicitly use transforms such as DCT or wavelets before quantization and entropy coding. Wavelet transforms are particularly relevant because JPEG~2000 uses biorthogonal wavelets and lifting to achieve perfect reconstruction and strong energy compaction~\cite{skodras2001jpeg2000,taubman2002jpeg2000}. Motivated by these properties, learned codecs have reintroduced frequency or wavelet structure at different stages. Learned lifting and wavelet-like transforms~\cite{Ma_2022_PAMI,sahin2023learnedDWT} provide invertible analysis transforms, while frequency-disentangled feature designs~\cite{Zafari_2023,pakdaman2024channelDecorr} encourage more structured latent representations. AuxT~\cite{Li2025AuxT} uses wavelet-based linear shortcuts to assist the optimization of nonlinear transforms. These approaches demonstrate that transform structure is useful in LIC, but they do not place a channel-wise lifting transform directly around the Q/K/V computation of spatial attention, nor do they construct a two-level channel-wise wavelet packet for slice-based entropy coding.

Several transformer-based codecs are closer to our setting because they introduce frequency or channel structure inside the transform. FAT~\cite{FAT} performs frequency-decomposition window attention and directional spatial modeling. BiSCT~\cite{soltani2024bilevelSCT} combines spatial high/low-frequency attention with channel-aware self-attention, and SCH~\cite{xu2024windowChannelWavelet} uses window-based channel attention together with DWT-based frequency-dependent downsampling. These designs confirm that frequency-aware and channel-aware modeling are important for learned compression. However, their mechanisms differ from ours: they either split spatial-frequency components, add explicit channel-attention branches, or use DWT mainly for sampling and receptive-field control. Our ChWDTA keeps the original windowed spatial tokenization and applies the wavelet transform along the channel axis before the Q/K/V projections, before transforming it back via the inverse wavelet transform. Thus the attention remains spatial attention, while its bilinear kernel is evaluated in a wavelet-transformed channel coordinate system.

The closest wavelet-domain LIC methods are WeConvene~\cite{Fu_EECV} and its 3D extension~\cite{Fu_3DM}. WeConvene inserts spatial DWT into convolutional layers and entropy coding; its entropy model first codes low-frequency spatial coefficients and then uses them to predict high-frequency coefficients. The 3D extension further applies multi-level spatial--channel DWT before convolution and entropy modeling. These methods provide strong evidence that wavelet-domain processing can improve learned compression. Nevertheless, their main operators are convolutional, and their entropy layouts are organized by spatial or 3D subbands. They do not analyze how a channel-wise wavelet transform changes the bilinear kernel of windowed spatial attention, nor do they use a two-level channel-wise wavelet packet whose four equal-sized subbands are further arranged as balanced ChARM-style entropy slices.

Note that the learned feature channels do not have the same physical frequency ordering as image pixels. Therefore the low/high branches in ChWDTB and ChWP are not claimed to be physical spatial-frequency bands. They are the two branches induced by wavelet transform, initialized from CDF~9/7 and optionally learned. This interpretation is important: the transform is a reversible, wavelet-structured channel reparameterization. Its benefit comes from reducing the mismatch of restricted attention and entropy models, not from discarding information.

In summary, our method uses channel-wise wavelets in two coordinated places. In ChWDTB, the lifting-based transform is an invertible wrapper around spatial attention: tokenization remains spatial, but the Q/K/V projections are computed from wavelet-transformed channel coordinates. In ChWP, a two-level wavelet packet organizes the latent into four equal-sized channel subbands; in our implementation, these subbands are further split into eight balanced sequential slices for entropy coding. This combination is different from pure transformer backbones, spatial frequency-aware attention, auxiliary wavelet shortcuts, and CNN-based wavelet-domain codecs.

\section{The Architecture of the Proposed Scheme}

Fig.~\ref{fig:framework} shows the architecture of the proposed LIC scheme. The input color image $x$ has dimensions $W \times H \times 3$, with pixel values normalized to $[-1,1]$. The analysis transform $g_a$ maps the input image to latent features, the synthesis transform $g_s$ reconstructs the image, and the hyperprior networks $h_a$ and $h_s$ use side information to enhance entropy modeling.

\begin{figure}[t]
  \centering
  \includegraphics[scale=0.60]{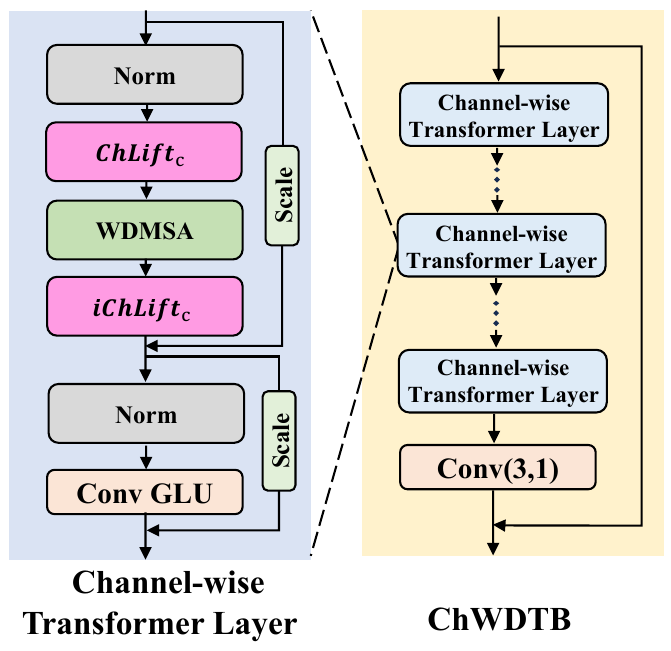}
  \caption{Details of the ChWDTB block. ChWDMSA denotes Channel-wise Wavelet-Domain Multi-head Self-Attention, while $\mathrm{WT}_c$ and $\mathrm{IWT}_c$ denote the channel-wise wavelet transform and its inverse, respectively. Each Channel-wise Wavelet-Domain Transformer Layer (ChWDTL) wraps channel-wise wavelet reparameterization around windowed spatial attention.}
  \label{fig:transformer}
\end{figure}

In existing hybrid CNN--transformer schemes~\cite{Liu_2023_CVPR,Feng2025LALIC,Lu_cvpr_2025}, the encoder/decoder networks include interleaved CNN layers and spatial transformer layers. In our scheme, we keep this efficient spatial-token backbone but replace each conventional transformer block with a channel-wise wavelet-domain transformer block (ChWDTB), whose details are shown in Fig.~\ref{fig:transformer}. The encoder therefore still consists of interleaved ResNet-based convolution layers (denoted as ResUnit in Fig.~\ref{fig:framework}) and transformer layers, but each transformer layer computes attention after a channel-wise wavelet reparameterization and maps the result back with the inverse transform. The same ChWDTB is also used in the hyperprior pathway and in the attention network of the channel-wise entropy-coding module.

\begin{figure*}[!t]
\centering
\subfloat[]{\includegraphics[width=0.32\linewidth]{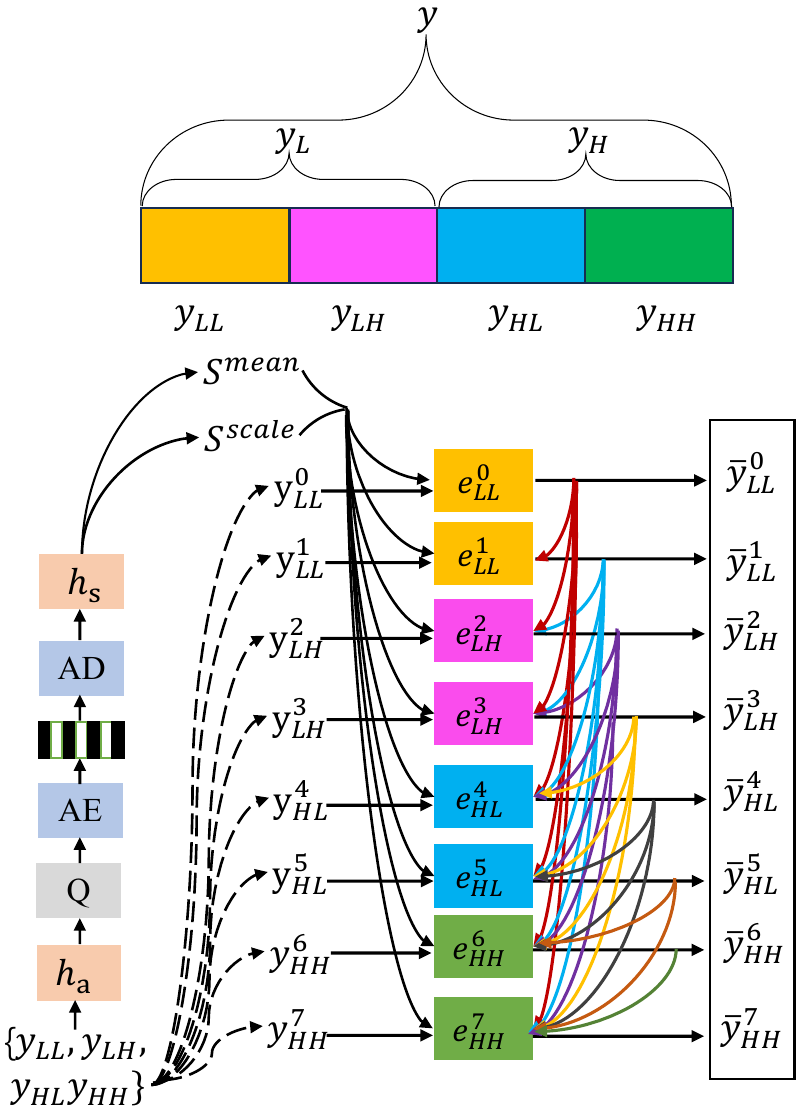}%
\label{fig:frameworkentropy}}
\hfil
\subfloat[]{\includegraphics[width=0.64\linewidth]{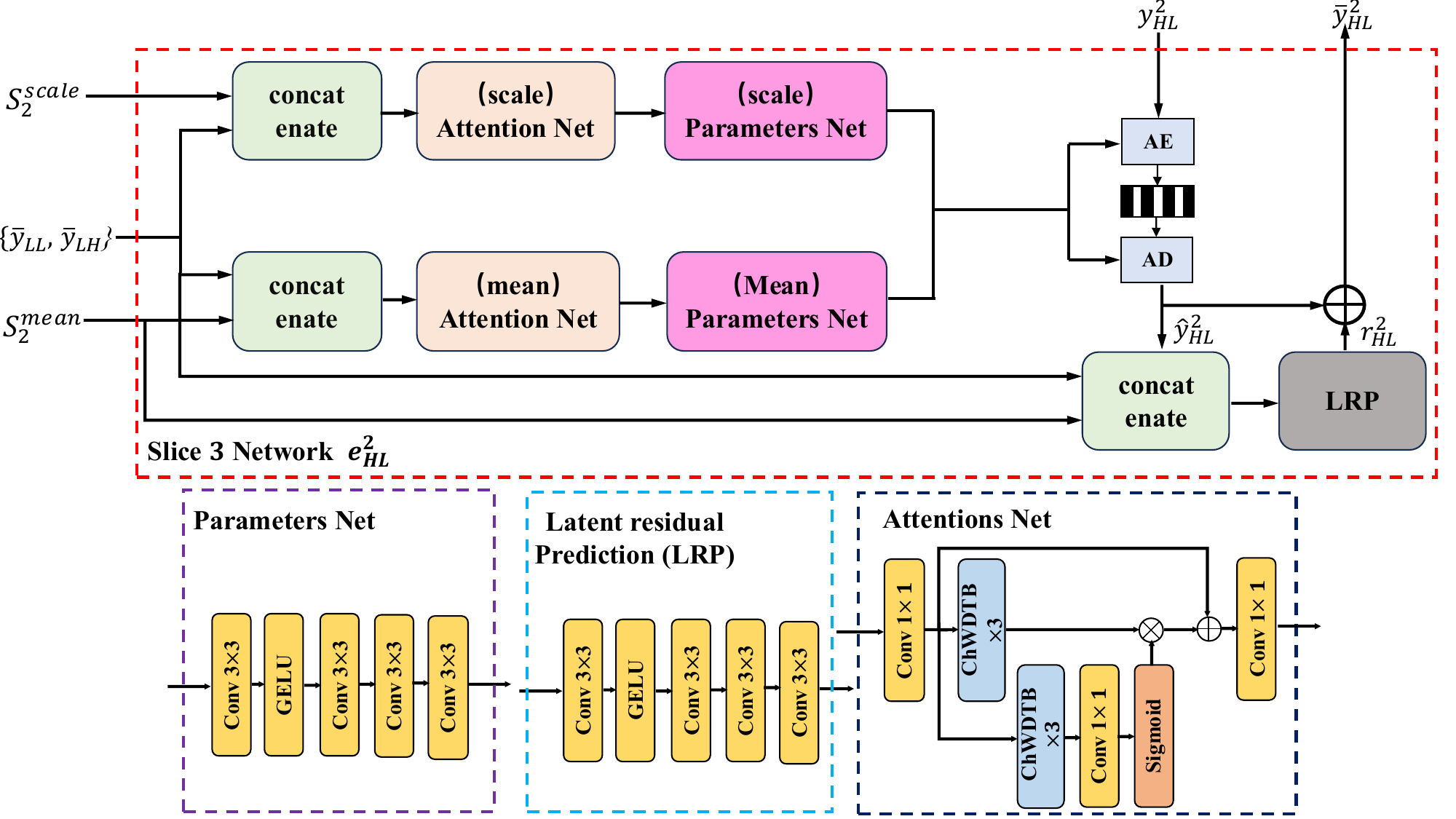}%
\label{fig:entropy_detailed_structure}}
\caption{The ChARM entropy-coding model. (a) After a two-level wavelet packet (WP) decomposition, the latent representation $y$ is partitioned into four equal-sized subbands. In the implementation, each subband is further split into two channel slices, producing $S=8$ sequential slices for entropy coding. (b) The coding process for each slice.}
\label{fig:charm_overall}
\end{figure*}

At the end of the encoder, we apply a channel-wise two-level wavelet packet transform (ChWP) to partition the channels into four equal-sized subbands~\cite{WavePacket}, as depicted in Fig.~\ref{fig:frameworkentropy}. This better aligns the latent representation with slice-based channel-wise entropy coding. 

The ChWP output subbands are quantized and entropy coded with arithmetic coding using the slice-based channel-wise autoregressive model (ChARM). In the default implementation, each of the four subbands is uniformly split into two channel slices, giving $S=8$ sequential slices. We refer to this setting as the \textbf{8-slice} configuration. ChARM does not use spatial context, so all spatial positions within each slice are processed in parallel, while the slices themselves are decoded sequentially. For completeness, we also evaluate a lightweight \textbf{4-slice} variant that directly uses the four ChWP subbands as four sequential slices without splitting each subband into two. Unless explicitly stated, the proposed method refers to the 8-slice setting.

The hyperprior network also uses ResUnit and ChWDTB. The extracted side information $\mathbf{z}$ is coded using the simple factorized entropy model (without using context).

\subsection{Channel-wise Wavelet-Domain Transformer Block (ChWDTB)}
\label{sec:chwdtb}

\textbf{Core idea.} ChWDTB does not change the spatial tokenization pattern of the attention mechanism itself. Instead, before computing Q/K/V projections, it applies an invertible 1D wavelet transform along the channel axis to each spatial token. After attention, the inverse channel-wise transform restores the original channel coordinate system. This simple wrapper changes the geometry of the attention kernel: the bilinear score between two tokens is now evaluated in a wavelet-transformed channel space.

Fig.~\ref{fig:transformer} shows the details of the ChWDTB block, which follows the overall transformer-block organization of~\cite{Lu_cvpr_2025} but changes the coordinate system in which spatial attention computes its Q/K/V projections. The standard windowed spatial self-attention is not replaced by channel attention. Instead, before the Q/K/V projections, each spatial token is reparameterized by an invertible one-dimensional wavelet transform along the channel axis; after attention, the corresponding inverse transform maps the result back to the original channel coordinates.

Let $\mathbf{F}_{\mathrm{in}}\in\mathbb{R}^{B\times C\times H'\times W'}$ be the normalized input of a Channel-wise Wavelet-Domain Transformer Layer (ChWDTL), where $B$ is the batch size, $C$ is the number of channels (even), and $H'\times W'$ is the spatial resolution. The core ChWDTA operation inside ChWDTL consists of four steps:

\begin{align}
\text{(Step 1)} &\quad (\mathbf{S},\mathbf{D}) = \mathrm{WT}_c(\mathbf{F}_{\mathrm{in}}), \nonumber\\
\text{(Step 2)} &\quad \mathbf{T} = \mathrm{concat}(\mathbf{S},\mathbf{D}),  \nonumber\\
\text{(Step 3)} &\quad \mathbf{A} = \mathrm{ChWDMSA}_{\mathrm{sp}}(\mathbf{T}),  \nonumber\\
\text{(Step 4)} &\quad \mathbf{F}_{\mathrm{out}} = \mathrm{IWT}_c(\mathbf{A}),
\label{eq:chwdtb_flow}
\end{align}
where $\mathrm{WT}_c$ and $\mathrm{IWT}_c$ represent the wavelet transform and inverse wavelet transform along the channel direction. $\mathbf{S}$ and $\mathbf{D}$ are the two channel branches produced by the wavelet transform (named ``smooth'' and ``detail'' by convention), and $\mathrm{ChWDMSA}_{\mathrm{sp}}(\cdot)$ denotes the windowed spatial multi-head self-attention computed after channel-wise wavelet reparameterization. The low/high terminology refers only to the two branches induced by the wavelet transform; it does not imply physical spatial-frequency bands in the image domain.

\subsubsection{Why Does the Wavelet Transform Help? A Covariance-Sparsification View}

To understand the effect of the channel-wise wavelet transform, we analyze the bilinear attention kernel. 

For one attention window, let $\f_i \in \mathbb{R}^C$ be the channel vector of the $i$-th spatial token. The standard attention score before softmax is:

\begin{equation}
\ell_{ij}^{\text{(orig)}} = \frac{(\W_Q \f_i)^\top (\W_K \f_j)}{\sqrt{d}} = \frac{\f_i^\top \M \f_j}{\sqrt{d}},
\label{eq:standard_bilinear_kernel}
\end{equation}
where $\M \triangleq \W_Q^\top \W_K$ is the learned attention kernel matrix, and $d$ is the head dimension.

In ChWDTA, we first transform each token to the channel-wise wavelet domain: $\uv_i = \W \f_i$, where $\W$ is the analysis matrix of the wavelet transform. The attention score becomes:

\begin{equation}
\ell_{ij}^{\text{(ChWDTA)}} = \frac{(\W_Q \uv_i)^\top (\W_K \uv_j)}{\sqrt{d}} = \frac{\f_i^\top \W^\top \M \W \f_j}{\sqrt{d}}.
\label{eq:wavelet_kernel}
\end{equation}

\textbf{Comparison.} Comparing Eq.~\eqref{eq:standard_bilinear_kernel} and Eq.~\eqref{eq:wavelet_kernel}, we see that the wavelet transform is not an inert preprocessing step: it changes the bilinear geometry in which spatial tokens are compared. The original attention operates on $\f$ with kernel $\M$, while ChWDTA operates on $\f$ with kernel $\W^\top \M \W$, which is equivalent to performing standard attention on the transformed representation $\uv$.

\subsubsection{Expected Attention Scores and Covariance Structure}

Why is this transformation beneficial? The answer lies in how the attention score interacts with the channel covariance structure. Let $\bm{\mu} = \E[\f_i]$ be the mean channel vector (assumed stationary across positions). The expected original attention score is:

\begin{equation}
\E[\ell_{ij}^{\text{(orig)}}] = \frac{1}{\sqrt{d}} \left( \Tr(\M \Sigm_{ji}) + \bm{\mu}^\top \M \bm{\mu} \right),
\label{eq:expected_attention_general}
\end{equation}

where $\Sigm_{ji} = \E[(\f_j - \bm{\mu})(\f_i - \bm{\mu})^\top]$ is the cross-position channel covariance. Under the common mean-field approximation that replaces $\Sigm_{ji}$ with the marginal channel covariance $\Sigm_f = \cov(\f)$ for all $i,j$ (including $i=j$), this simplifies to:

\begin{equation}
\E[\ell_{ij}^{\text{(orig)}}] \approx \frac{1}{\sqrt{d}} \left( \Tr(\M \Sigm_f) + \bm{\mu}^\top \M \bm{\mu} \right).
\label{eq:expected_attention_meanfield}
\end{equation}

Note that Eq.~\eqref{eq:expected_attention_meanfield} is an approximation by assuming that cross-position covariances are identical to the marginal covariance.

In the wavelet domain, the same expression becomes:

\begin{equation}
\E[\ell_{ij}^{\text{(ChWDTA)}}] \approx \frac{1}{\sqrt{d}} \left( \Tr(\M_u \Sigm_u) + \bm{\mu}_u^\top \M_u \bm{\mu}_u \right),
\label{eq:expected_wavelet_meanfield}
\end{equation}

where $\Sigm_u = \W \Sigm_f \W^\top$, $\bm{\mu}_u = \W \bm{\mu}$, and $\M_u$ is the attention kernel learned in the wavelet domain.

\subsubsection{Covariance Sparsification}

The key observation is that a well-designed wavelet transform can sparsify the channel covariance matrix. Writing $\Sigm_u$ in block form:

\begin{equation}
\Sigm_u = \W \Sigm_f \W^\top = \begin{bmatrix}
\Sigm_{LL} & \bm{\Delta}_{LH} \\
\bm{\Delta}_{HL} & \Sigm_{HH}
\end{bmatrix},
\label{eq:covariance_block}
\end{equation}

where $\Sigm_{LL}$ and $\Sigm_{HH}$ capture within-branch covariances, while $\bm{\Delta}_{LH}$ and $\bm{\Delta}_{HL}$ capture cross-branch couplings. We define the \emph{cross-branch coupling ratio}:

\begin{equation}
r_{\text{off}} = \frac{\|\bm{\Delta}_{LH}\|_F + \|\bm{\Delta}_{HL}\|_F}{\|\Sigm_{LL}\|_F + \|\Sigm_{HH}\|_F}.
\label{eq:roff_definition}
\end{equation}

A smaller $r_{\text{off}}$ means the wavelet representation has weaker coupling between the two channel groups, i.e., a more block-sparse covariance structure.

\textbf{Why this matters for attention.} Consider the contribution of cross-branch interactions to an attention score. By Cauchy-Schwarz inequality, for any matrices $\M_{LH}$ and $\bm{\Delta}_{HL}$ of compatible sizes:

\begin{equation}
|\Tr(\M_{LH} \bm{\Delta}_{HL})| \leq \|\M_{LH}\|_F \|\bm{\Delta}_{HL}\|_F.
\label{eq:cross_bound}
\end{equation}

Therefore, reducing the cross-covariance norm $\|\bm{\Delta}_{HL}\|_F$ directly reduces the maximum possible contribution from cross-branch interactions that the finite-capacity attention head must model. This provides a clean explanation: the wavelet transform serves as a structured preconditioner that aligns the channel representation with the inductive bias of spatial attention.

\subsection{Channel-wise Wavelet Packet-based Entropy Coding}
\label{sec:entropy}

The entropy-coding module is shown in Fig.~\ref{fig:charm_overall}. Before entropy coding, we apply a two-level Channel-wise Wavelet Packet (ChWP) transform~\cite{WavePacket}:

\begin{align}
\text{(Level 1)} &\quad (\mathbf{L}_{1},\mathbf{H}_{1}) = \mathrm{WT}_c(\mathbf{Y}), \nonumber\\
\text{(Level 2)} &\quad (\mathbf{LL},\mathbf{LH}) = \mathrm{WT}_c(\mathbf{L}_{1}), \nonumber\\
\text{(Level 2)} &\quad (\mathbf{HL},\mathbf{HH}) = \mathrm{WT}_c(\mathbf{H}_{1}).
\label{eq:chwp_decomp}
\end{align}

This produces four equal-sized channel subbands: $\mathbf{LL}$, $\mathbf{LH}$, $\mathbf{HL}$, $\mathbf{HH}$. This balanced layout is naturally aligned with slice-based entropy coding, which usually uses slices of the same size. In the default 8-slice layout, each subband is split into two channel slices, so $S=8$ slices are decoded sequentially. The 4-slice variant directly uses the four ChWP subbands as the four sequential slices, which has lower complexity.

\textbf{Why ChWP helps entropy coding.} The compression role of ChWP follows the same covariance-sparsification view as ChWDTB. After the ChWP transform, we group the latent into ordered entropy slices. Let
\begin{equation}
\Sigm^{w} = \cov(\mathbf{y}^{w} \mid \hat{\mathbf{z}}) = [\Sigm^{w}_{ij}]_{i,j=1}^{S}
\label{eq:chwp_cov}
\end{equation}
be the conditional covariance of the ChWP-domain latent grouped into $S$ slices, where block $\Sigm^{w}_{ii}$ is the within-slice covariance and $\Sigm^{w}_{ij}$ ($i \neq j$) is the cross-slice covariance. A compact measure of how well the slice ordering aligns with the entropy model is:

\begin{equation}
r_{\text{ent}} = \frac{\sum_{i<j} \|\Sigm^{w}_{ij}\|_F + \sum_{i=1}^{S} \|\text{offdiag}(\Sigm^{w}_{ii})\|_F}{\|\Sigm^{w}\|_F}.
\label{eq:entropy_sparsity}
\end{equation}

The first term measures residual cross-slice coupling, while the second term measures within-slice off-diagonal covariance that a diagonal likelihood model cannot explicitly represent. By reducing $r_{\text{ent}}$, ChWP makes the conditional distribution easier for the restricted slice-wise model to approximate.

\textbf{The entropy coding model.} Conditioned on the decoded hyperprior $\hat{\mathbf{z}}$, ChARM codes the slices sequentially with a learned model $q_i(\hat{\mathbf{y}}^{w}_{i} \mid \hat{\mathbf{y}}^{w}_{<i}, \hat{\mathbf{z}})$. The expected arithmetic-coding rate is:

\begin{equation}
\bar{R} = \sum_{i=1}^{S} \E[-\log_2 q_i(\hat{\mathbf{y}}^{w}_{i} \mid \hat{\mathbf{y}}^{w}_{<i}, \hat{\mathbf{z}})].
\label{eq:cross_entropy_rate}
\end{equation}

For the $i$-th slice, the entropy model is diagonal after conditioning on previous slices and the hyperprior:

\begin{equation}
q_i(\hat{\mathbf{y}}^{w}_{i} \mid \hat{\mathbf{y}}^{w}_{<i}, \hat{\mathbf{z}}) = \prod_{j=1}^{d_i} q_{ij}(\hat{y}^{w}_{ij} \mid \hat{\mathbf{y}}^{w}_{<i}, \hat{\mathbf{z}}),
\label{eq:slice_factorized}
\end{equation}
where $d_i$ is the number of coefficients in the slice. For each coefficient, we use a discretized Gaussian:

\begin{equation}
q_{ij}(\hat{y}^{w}_{ij} \mid \hat{\mathbf{y}}^{w}_{<i}, \hat{\mathbf{z}}) = \Phi\!\left(\frac{\hat{y}^{w}_{ij} + \frac{1}{2} - \mu_{ij}}{\sigma_{ij}}\right) - \Phi\!\left(\frac{\hat{y}^{w}_{ij} - \frac{1}{2} - \mu_{ij}}{\sigma_{ij}}\right),
\label{eq:discrete_gaussian}
\end{equation}
where $\Phi(\cdot)$ is the standard Gaussian CDF.

\textbf{Connection to rate savings.} Under a local Gaussian approximation, the rate penalty of using a diagonal model instead of the true conditional covariance can be quantified as:

\begin{equation}
\Delta R_i \approx \frac{1}{2\ln 2} \log \frac{\det(\operatorname{diag}(\Sigm_{i\mid <i,\hat{z}}))}{\det(\Sigm_{i\mid <i,\hat{z}})}.
\label{eq:diag_gap}
\end{equation}

Eq.~\eqref{eq:diag_gap} is used only as an interpretive diagnostic, not as an exact rate prediction. It indicates that when the conditional covariance becomes closer to diagonal, the diagonal entropy model incurs less mismatch. Therefore, a ChWP ordering with weaker cross-slice and within-slice off-diagonal covariance is better matched to the entropy model.

Fig.~\ref{fig:charm_overall}(b) shows the coding process for each slice $e_{\text{Slice}}$, which is based on~\cite{channel,Liu_2023_CVPR,Fu_EECV}. It takes as input the scale and mean parameters produced by the hyperprior decoder, together with the previously decoded slices. These inputs are processed by the AttentionNet and the parameter networks to infer the distribution parameters of the current slice. In addition, we adopt the Latent Residual Prediction (LRP) module from~\cite{channel} to estimate the quantization error, which is then added to the dequantized latent to obtain the final reconstruction. Compared with the residual-block design used in previous work~\cite{Fu_EECV}, our AttentionNet replaces those residual blocks with the proposed ChWDTB blocks to improve entropy-parameter estimation.

\subsection{Learnable Lifting Wavelet}
\label{sec:lifting}

The wavelet transform used in ChWDTB and ChWP is implemented by lifting~\cite{sweldens1996lifting,taubman2002jpeg2000}. We adopt a CDF~9/7-style topology with four predict/update steps and a paired scaling. For each spatial location, let $\mathbf{s}$ and $\mathbf{d}$ be the even and odd channel sequences, and let $\mathcal{A}_{s}$ and $\mathcal{A}_{d}$ denote the circular neighbor-sum operators used by the lifting filters. The learnable lifting template is:

\begin{align}
\mathbf{d} &\leftarrow \mathbf{d} + \alpha \mathcal{A}_{s} \mathbf{s}, \quad
\mathbf{s} \leftarrow \mathbf{s} + \beta \mathcal{A}_{d} \mathbf{d}, \nonumber\\
\mathbf{d} &\leftarrow \mathbf{d} + \gamma \mathcal{A}_{s} \mathbf{s}, \quad
\mathbf{s} \leftarrow \mathbf{s} + \delta \mathcal{A}_{d} \mathbf{d}, \nonumber\\
\mathbf{s} &\leftarrow K \mathbf{s}, \quad
\mathbf{d} \leftarrow K^{-1} \mathbf{d}.
\label{eq:lifting_steps}
\end{align}

The parameters $(\alpha,\beta,\gamma,\delta,K)$ are initialized from the CDF~9/7 wavelet, and $K$ is parameterized as $K = \exp(\kappa)$ to keep the scaling nonzero. Each predict/update step is invertible by subtracting the same update, and the paired scaling is invertible by reciprocal scaling. Therefore, the inverse channel-wise wavelet transform is obtained by reversing the order and changing the signs of the predict/update coefficients.




\subsection{Complexity Analysis}

We analyze the complexity of the main components as follows.

\paragraph{Attention Path}
For tokens of shape $B \times H' \times W' \times C$ and window size $p$, windowed MSA has time complexity $\mathcal{O}(B H' W' p^{2} C)$ and activation memory $\mathcal{O}(B H' W' C)$.
Adding $\mathrm{WT}_c$ before attention and $\mathrm{IWT}_c$ after attention costs $\mathcal{O}(B H' W' C)$ operations and negligible extra memory, which is small for typical $p$.

\paragraph{Entropy Coding}
Pixel-wise autoregressive models are inherently sequential and scale with the number of spatial positions. Our ChARM entropy pipeline has only $S$ sequential slice steps, typically $S=8$, and each step performs a fully parallel arithmetic decode over all $H' \times W'$ positions.

The channel-wise two-level wavelet packet has linear complexity $\mathcal{O}(B H' W' C)$, so the overall latency is low.

\paragraph{Parameter and Run Time Footprint}
If a fixed $\mathrm{WT}_c$ is used (e.g., CDF~9/7), no learnable parameter is needed. The learnable lifting wavelet adds only a few parameters: $\alpha,\beta,\gamma,\delta,K$. The feature map sizes are unchanged, so activation memory and bandwidth usage remain the same as the spatial transformer.

\subsection{Implementation Details}

All wavelet modules are initialized from the CDF~9/7 lifting coefficients with periodic boundary extension \cite{taubman2002jpeg2000}. Depending on the experiment, a module can either keep this initialization fixed or fine-tune its own lifting parameters independently. Since each lifting transform contributes only five scalar parameters, allowing module-specific learning adds negligible parameter overhead.

The hyperprior follows a standard design, and the conditional Gaussian is parameterized by $h_s$ together with lightweight cross-channel predictors that operate on previously decoded slices with bounded support.

Training is performed independently for each rate--distortion multiplier. We use Adam and the three-stage schedule specified in Sec.~\ref{experiments}; no cosine learning-rate decay is used.

Unless stated otherwise, the two-level wavelet packet is enabled in the entropy coding and the wavelet-domain attention is active in all transformer blocks.

\section{Experiments}
\label{experiments}

We follow the experimental settings of recent learned methods~\cite{Zou_2022_CVPR,Lu_cvpr_2025} and train our models on an OpenImages subset containing about 300k images~\cite{Krasin2016openimage}. Unless noted otherwise, we minimize an MSE-based R--D objective with $\lambda\in\{0.0025,0.0035,0.0067,0.0130,0.0250,0.0500\}$. For MS-SSIM, we optimize our models with $\lambda\in\{3,5,8,16,32,64\}$. Each model is optimized with Adam using a three-stage schedule. In the first two stages, training uses random $256\times256$ crops with a batch size of 8: epochs 1--45 use a learning rate of $10^{-4}$, and epochs 46--65 continue from the same checkpoint with a learning rate of $10^{-5}$. In the final stage, the training crop size is increased to $512\times512$, and the model is fine-tuned for another 5 epochs with the learning rate kept at $10^{-5}$. Thus, the complete training schedule contains 70 epochs, and the batch size is kept at 8 in all three stages.

Evaluation follows common practice on \textbf{Kodak} (24 images, $768\times512$) \cite{Kodak}, \textbf{Tecnick} (100 images, $1200\times1200$) \cite{Tecnick}, and \textbf{CLIC Professional Validation} (41 images, around $2048\times1440$) \cite{CLIC_test_2021}. We report bits per pixel (bpp) versus \textbf{PSNR} and \textbf{MS-SSIM} \cite{MS-SSIM}, and report BD-rate (\%) with respect to our VVC-Intra 4:4:4 anchor generated by VTM~9.1. BD-rate is computed from the R--D curves over the common bitrate range of the compared methods. Encoding and decoding time is measured end to end on an NVIDIA RTX 4090 GPU with batch size 1 at $768\times512$ input resolution.

We compare with the strongest classical image codec VVC and recent learned methods including \textbf{ELIC}~\cite{He_2022_CVPR}, \textbf{STF}~\cite{Zou_2022_CVPR}, \textbf{TCM}~\cite{Liu_2023_CVPR}, \textbf{FAT}~\cite{FAT}, \textbf{MLIC++}~\cite{jiang2023mlicpp}, \textbf{WeConvene}~\cite{Fu_EECV}, \textbf{LALIC}~\cite{Feng2025LALIC}, and \textbf{DCAE}~\cite{Lu_cvpr_2025}. For fair comparison, we reuse their public models when available and adopt their evaluation protocols. 

The analysis/synthesis backbones follow Swin-style windowed attention with residual depthwise-convolutional feed-forward and residual scaling. The default entropy path uses two-level channel-wise wavelet packet decomposition and $S=8$ channel slices, obtained by splitting each of the four ChWP subbands into two equal channel slices. This default setting is denoted as the 8-slice layout. We additionally report a 4-slice variant, where the four ChWP subbands are coded directly without the final two-way split inside each subband, to show the hardware-oriented complexity trade-off. Unless otherwise specified, every wavelet module is initialized from CDF~9/7. In the more flexible setting, different modules may learn their own lifting coefficients rather than sharing one global wavelet, which is consistent with the module-specific interpretation discussed in Sec.~\ref{sec:lifting}.

Following \cite{Lu_cvpr_2025}, we set the numbers of transformer layers in the Transformer Block at three scales as $(T_1,T_2,T_3)=(1,2,12)$. The hyperprior module uses a single transformer layer. For channel capacities, we choose $(C_1,C_2,C_3,C_4,C_5)=(96,144,256,192,192)$. The latent representation $\mathbf{y}$ has dimension $320$ and the side information $\mathbf{z}$ has dimension $192$. The attention head counts for transformer layers in the encoder $g_a$ and decoder $g_s$ are ${8,16,32,32,16,8}$. The attention head count in the hyperprior encoder $h_a$ and hyperprior decoder $h_s$ is $32$. Window sizes are $8{\times}8$ for $g_a$ and $g_s$, and $4{\times}4$ for the hyperprior module.

Unless otherwise stated, all ablations use the same training schedule, datasets, and evaluation protocol as the full model, and change only the component explicitly mentioned in each table. This avoids conflating the effect of the wavelet modules with unrelated optimization or backbone variations.

\subsection{Trade-off between Compression Efficiency and Computational Complexity}
\label{sec:speed}

\begin{table*}[!t]
\centering
\setlength{\tabcolsep}{2.5pt}
\fontsize{8.5pt}{11}\selectfont
\caption{Compression performance and complexity comparison. VVC VTM-9.1 is used as the anchor for BD-rate calculation. Boldface marks the best value in each column; for BD-rate, a more negative value is better.}
\label{rd_time}
\vspace{0.8em}
\begin{threeparttable}
\begin{tabular}{lccccccc}
\hline
\multirow{2}{*}{Model} & \multirow{2}{*}{\begin{tabular}[c]{@{}c@{}}Enc. Time$^\dagger$ \\ (ms)\end{tabular}} & \multirow{2}{*}{\begin{tabular}[c]{@{}c@{}}Dec. Time$^\dagger$ \\ (ms)\end{tabular}} & \multirow{2}{*}{\begin{tabular}[c]{@{}c@{}}kMACs\\ /pixel\end{tabular}} & \multirow{2}{*}{\begin{tabular}[c]{@{}c@{}}Params\\ (M)\end{tabular}} & \multicolumn{3}{c}{BD-rate} \\ \cline{6-8} 
                       & & & & & Kodak & CLIC Pro Valid & Tecnick \\ \hline
ELIC (CVPR'22) \cite{He_2022_CVPR}        & 97   & 103  & 573.88  & \textbf{36.93} & -7.88\%  & -5.12\%  & -8.63\%  \\
STF (CVPR'22) \cite{Zou_2022_CVPR}         & 106  & 107  & \textbf{511.17} & 99.86 & -4.13\%  & -0.98\%  & -2.44\%  \\
TCM (CVPR'23) \cite{Liu_2023_CVPR}         & 114  & 126  & 1823.58 & 76.57 & -11.73\% & -9.14\%  & -10.93\% \\
MLIC++ (NCW ICML'23) \cite{jiang2023mlicpp}  & 360  & 408  & 1282.81 & 116.72 & -15.02\% & -14.45\% & -17.21\% \\
FAT (ICLR'24) \cite{FAT}        & $>$1000 & $>$1000 & 1096.04 & 70.96 & -14.56\% & -10.79\% & -14.40\% \\
WeConvene (ECCV'24) \cite{Fu_EECV}   & 222  & 223  & 2343.11 & 107.15 & -8.35\%  & -6.88\%  & -7.53\%  \\
LALIC (CVPR'25) \cite{Feng2025LALIC}       & 189.0 & \textbf{95.4} & 667.26  & 66.13 & -15.26\% & -15.41\% & -17.63\% \\
DCAE (CVPR'25) \cite{Lu_cvpr_2025}        & 87   & 97   & 937.09  & 119.4  & -16.98\% & -16.98\% & -20.13\% \\ \hline
Ours (4-slice)  & \textbf{86} & 101  & 913.65  & 113.0  & -17.24\% & -18.42\% & -21.71\% \\ 
Ours (8-slice, default)  & 102  & 123  & 1204.68 & 189.1  & \textbf{-17.82\%} & \textbf{-19.15\%} & \textbf{-22.56\%} \\ \hline
\end{tabular}
\begin{tablenotes}[flushleft]
\footnotesize
\item[$\dagger$] End-to-end encoding/decoding latency is measured with batch size 1 on an NVIDIA RTX~4090 GPU at $768\times512$ input resolution.
\end{tablenotes}
\end{threeparttable}
\end{table*}

Table~\ref{rd_time} compares compression efficiency and complexity across representative learned codecs. BD-rate is reported with respect to our VVC-Intra 4:4:4 anchor generated by VTM~9.1 whenever the corresponding R--D points are available. For methods whose public models or complete R--D points are not available, we follow the numbers reported by the original papers. The number of parameters and kMACs/pixel are computed with the PyTorch Flops Profiler\footnote{\url{https://www.deepspeed.ai/tutorials/flops-profiler}} when public models are available; otherwise we follow the numbers reported by the original papers. This convention keeps the comparison reproducible while avoiding unverified re-implementation of methods without complete public checkpoints.

FAT~\cite{FAT} employs serial spatial context-adaptive entropy models to achieve superior coding performance; however, this significantly increases encoding/decoding time. In contrast, recent LIC methods, including ELIC~\cite{He_2022_CVPR}, TCM~\cite{Liu_2023_CVPR}, and WeConvene~\cite{Fu_EECV}, adopt GPU-friendly parallelizable entropy models that substantially reduce decoding time while maintaining comparable compression performance.

The default 8-slice model obtains the most favorable BD-rate values in Table~\ref{rd_time}: $-17.82\%$, $-19.15\%$, and $-22.56\%$ on Kodak, CLIC Professional Validation, and Tecnick, respectively. Relative to DCAE~\cite{Lu_cvpr_2025}, the corresponding margins are $0.84$, $2.17$, and $2.43$ percentage points. The larger margins on CLIC Professional Validation and Tecnick suggest that the proposed channel-wise wavelet-domain sparsification is particularly useful for higher-resolution images. The 8-slice model is therefore used as the default configuration for compression efficiency. The 4-slice model provides a lower-complexity operating point: compared with the 8-slice model, it reduces the parameter count from $189.1$M to $113.0$M and the kMACs/pixel from $1204.68$ to $913.65$, while still obtaining BD-rate values of $-17.24\%$, $-18.42\%$, and $-21.71\%$. In the low-latency region, the 4-slice model also has a lower encoding time, fewer kMACs/pixel, and better BD-rate than DCAE, with a comparable total latency on Kodak (187~ms versus 184~ms). Fig.~\ref{fig:latency_bdrate} visualizes this trade-off between latency and Kodak BD-rate.

\begin{figure*}[t]
\centering
\subfloat[PSNR]{\includegraphics[width=0.49\linewidth]{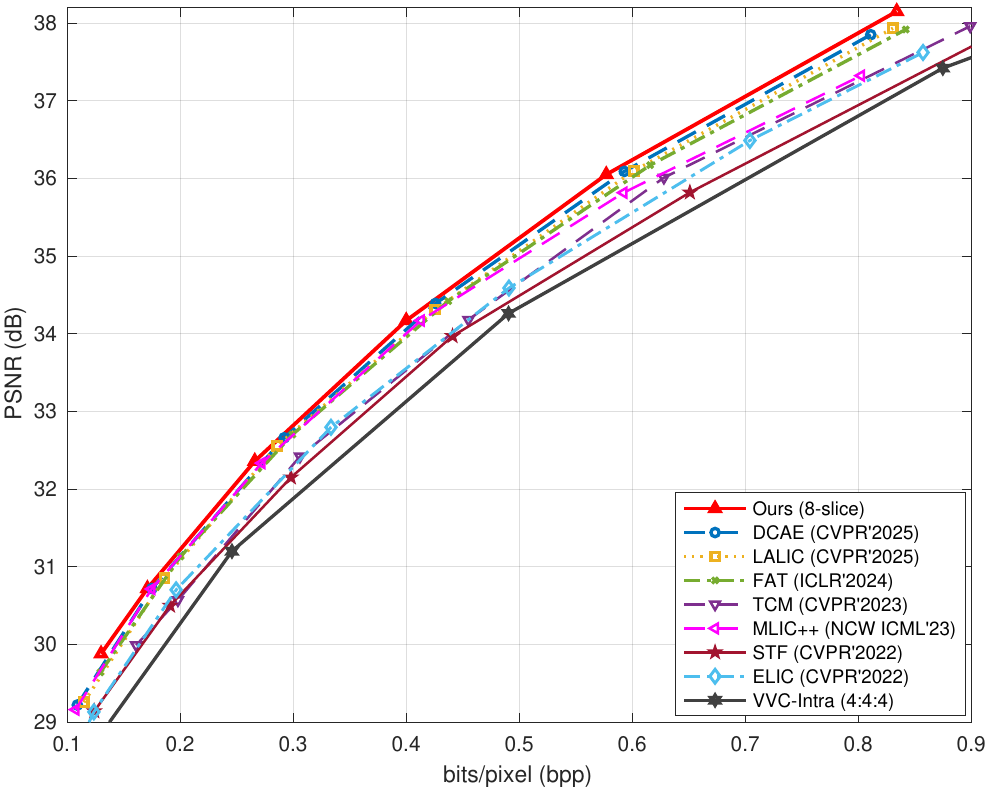}%
\label{fig:kodak_psnr}}
\hfil
\subfloat[MS-SSIM]{\includegraphics[width=0.49\linewidth]{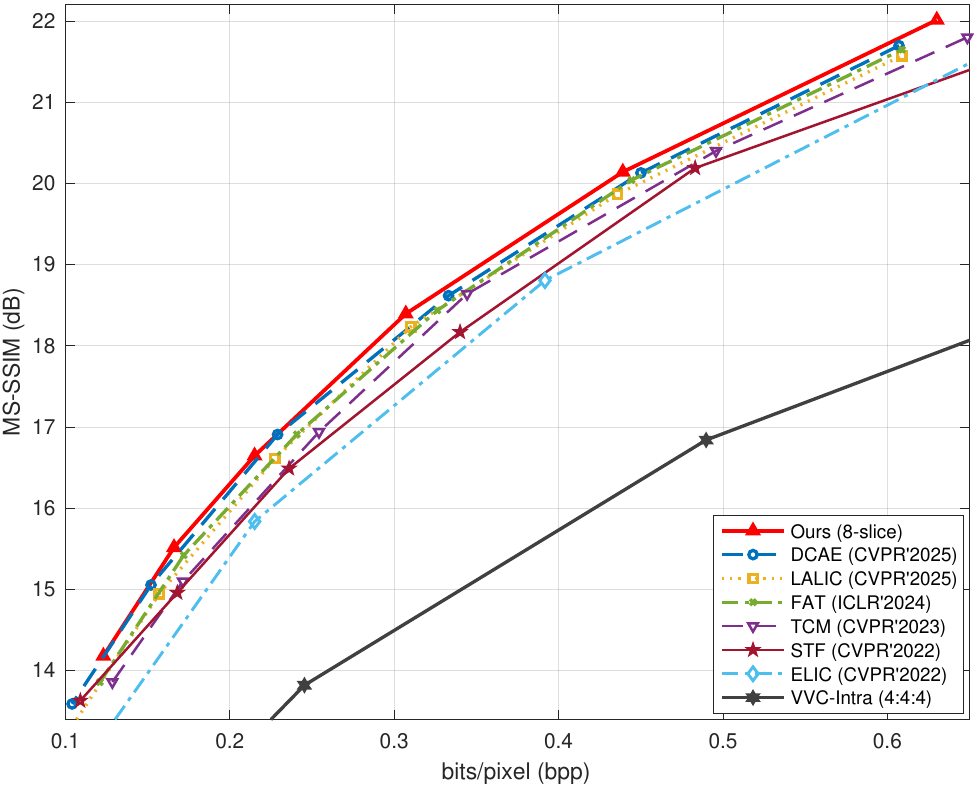}%
\label{fig:kodak_msssim}}
\caption{Evaluation results of different methods on the Kodak dataset in terms of PSNR and MS-SSIM.}
\label{fig:kodak_metrics}
\end{figure*}

\begin{figure*}[t]
\centering
\subfloat[Tecnick]{\includegraphics[width=0.49\linewidth]{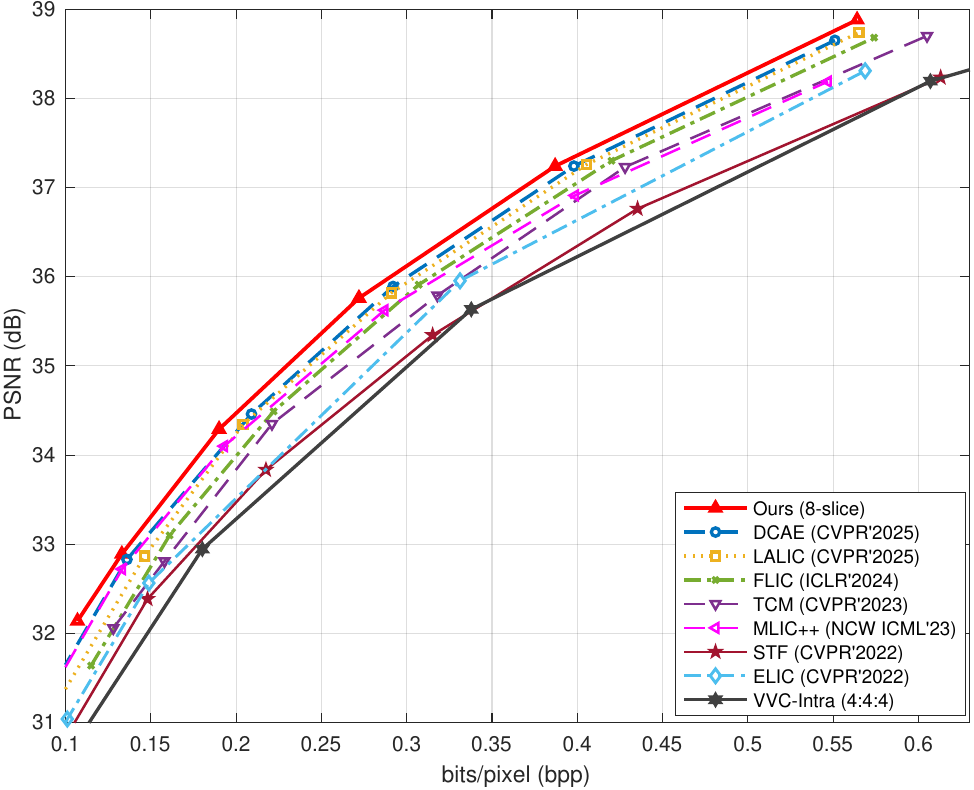}%
\label{fig:tecnick_psnr}}
\hfil
\subfloat[CLIC Pro]{\includegraphics[width=0.49\linewidth]{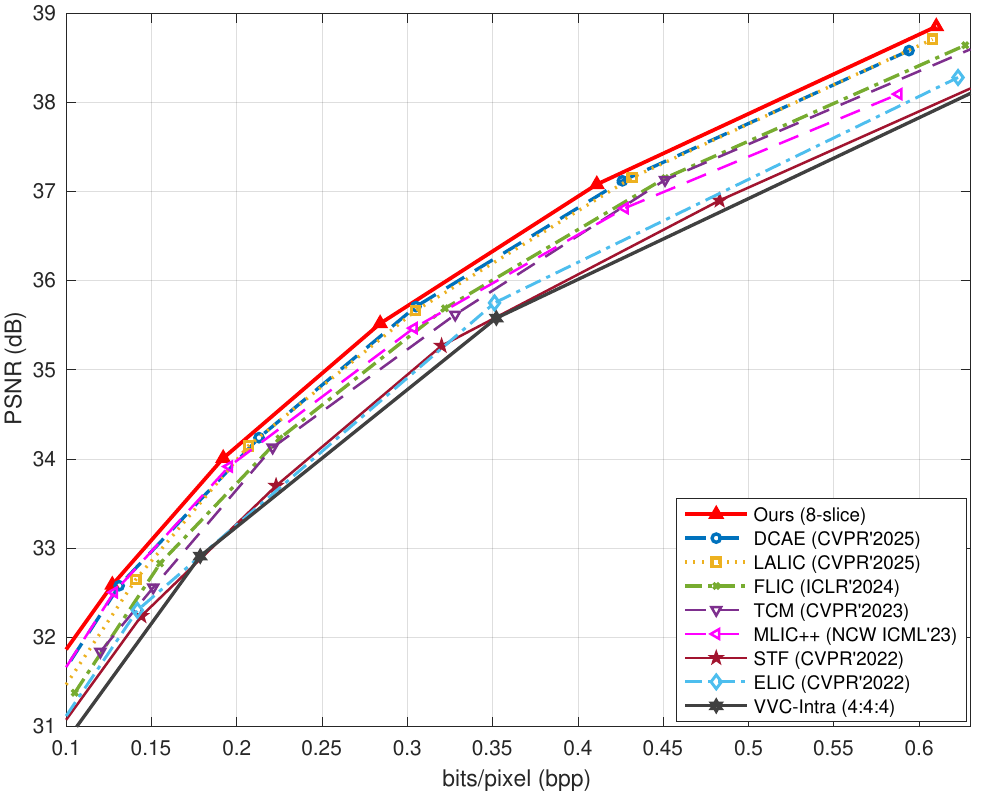}%
\label{fig:clic_pro_psnr}}
\caption{Evaluation results of different methods on the Tecnick and CLIC Professional Validation datasets in terms of PSNR. }
\label{fig:clic_tecnick_psnr}
\end{figure*}

\subsection{R--D Performance}

For clarity, the main R--D curves in Figs.~\ref{fig:kodak_metrics} and~\ref{fig:clic_tecnick_psnr} plot only the default 8-slice configuration. The 4-slice model is a lower-complexity entropy grouping variant rather than the default codec, and its R--D curve is very close to that of the 8-slice model, which makes the figure less readable if both curves are shown together. It is therefore summarized through BD-rate and complexity in Table~\ref{rd_time}, the rate--complexity comparison in Fig.~\ref{fig:latency_bdrate}, and the grouping ablation in Table~\ref{tab:ablate_depth}. This keeps the main R--D figures focused on comparison with prior codecs while still reporting the 4-slice complexity trade-off.

Fig.~\ref{fig:kodak_metrics} compares the average R--D performance of different methods on the Kodak dataset in terms of PSNR and MS-SSIM. In the PSNR results, DCAE~\cite{Lu_cvpr_2025} is the strongest competing method, while our scheme achieves better results over a wide range of bit rates. When the bit rate is below $0.3$ bpp, our method achieves slightly better performance than DCAE; when the bit rate is higher than $0.5$ bpp, the improvement becomes more visible, with about $0.15$--$0.2$ dB gain at high rates. Our method also outperforms other learned codecs and VVC. For MS-SSIM, our scheme also achieves better results than the compared learned codecs and VVC across the bitrate range.

Fig.~\ref{fig:tecnick_psnr} compares the PSNR performance of different methods on the Tecnick dataset. Among the competing methods, DCAE~\cite{Lu_cvpr_2025} achieves the best overall performance. Our method still achieves better results than DCAE at both low and high bit rates. When the bit rate is less than $0.2$ bpp, our scheme obtains slightly higher PSNR, while at medium and high bit rates it keeps a clear advantage over DCAE, LALIC~\cite{Feng2025LALIC}, and the other learned codecs. It also consistently outperforms VVC over a wide range of bit rates.

Fig.~\ref{fig:clic_pro_psnr} shows the PSNR results on the CLIC Professional Validation dataset, whose images have much higher resolution than Kodak. DCAE and LALIC are the closest competing methods, but our method achieves better R--D performance across the bitrate range. At low bit rates, our method is slightly better than the strongest baseline, and when the bit rate is higher than $0.4$ bpp, the gain becomes more obvious. Together with the Tecnick results, this shows that our method is particularly effective on high-resolution images, where richer spatial structures and stronger channel dependencies make the attention and entropy models benefit more from covariance sparsification.

\subsection{Visualization and Interpretability}

To support the interpretation in Sec.~\ref{sec:chwdtb}, we provide a compact channel-correlation visualization for a representative ChWDMSA block. The visualization is not used as a separate contribution; it is meant to show whether $\mathrm{WT}_c$ actually reorganizes the feature covariance in the way assumed by the second-order analysis. The protocol is deterministic: we hook the selected ChWDMSA block, compute channel-token correlations before and after $\mathrm{WT}_c$, rank channels by a fixed hybrid score that combines pre-$\mathrm{WT}_c$ dependency strength and per-channel entropy, and display the top-$48$ channels. In addition to the plotted correlation matrices, we compute the covariance-based cross-branch coupling ratio $r_{\text{off}}$ in Eq.~\eqref{eq:roff_definition} on the same selected channels as a scalar diagnostic. This avoids arbitrary channel selection, keeps the visualization reproducible, and directly connects the figure with the covariance-sparsification measure defined in Sec.~\ref{sec:chwdtb}.

\begin{figure*}[t]
  \centering
  \includegraphics[width=\textwidth]{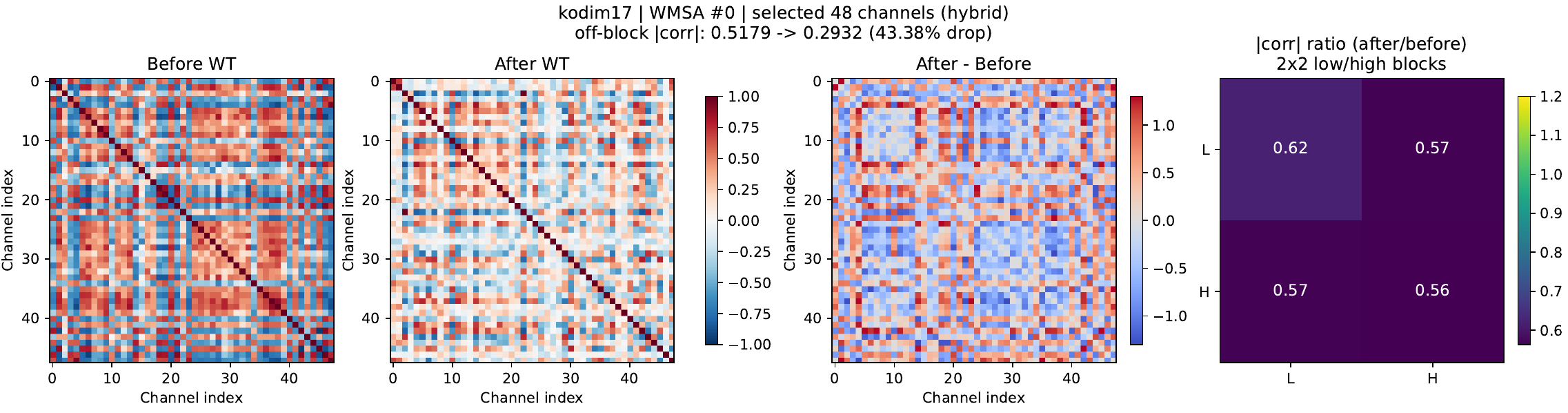}
  \caption{Representative channel-correlation visualization for ChWDTB at ChWDMSA~\#0 on a test image. The first two panels show the selected channel-correlation matrices before and after $\mathrm{WT}_c$, the third panel shows the element-wise difference, and the fourth panel reports the ratio of mean absolute correlation after/before within and across the two wavelet-induced channel groups. Channels are selected by the fixed hybrid ranking described in the text. In this example, the off-block mean absolute correlation decreases from $0.5179$ to $0.2932$, i.e., a $43.38\%$ drop. The $2\times2$ branch-wise block ratios are all below one $(0.62, 0.57; 0.57, 0.56)$, indicating that both within-group and cross-group correlations are weakened after the transform. The covariance-based cross-branch coupling ratio in Eq.~\eqref{eq:roff_definition}, computed on the same selected channels but not plotted as an additional panel, decreases from $0.6646$ to $0.5054$, corresponding to a $23.96\%$ reduction.}
  \label{fig:wt_corr_heatmap}
\end{figure*}

Fig.~\ref{fig:wt_corr_heatmap} provides empirical support for the sparsification argument in Sec.~\ref{sec:chwdtb}. The transform does not produce exact diagonalization; instead, it reorganizes a diffusely correlated channel basis into a more structured two-branch layout before the subsequent Q/K projections. The first three panels show that strong positive and negative channel correlations become visibly weaker after $\mathrm{WT}_c$. The fourth panel further summarizes this change at the block level: the smooth--smooth, smooth--detail, detail--smooth, and detail--detail mean-absolute-correlation ratios are all smaller than one. Therefore, the reduction is not limited to a single off-diagonal region; it is observed in both the within-branch and cross-branch blocks.

We also report the scalar coupling measure defined in Eq.~\eqref{eq:roff_definition}. For this representative example, $r_{\text{off}}$ decreases from $0.6646$ to $0.5054$, which is a $23.96\%$ reduction. This covariance-based number is more directly aligned with the theoretical definition than the plotted correlation heatmaps, while the heatmaps make the same phenomenon visually interpretable. Combined with the bound in Eq.~\eqref{eq:cross_bound}, the decrease in $r_{\text{off}}$ indicates that the cross-branch covariance terms available to the finite-capacity attention projection are reduced after $\mathrm{WT}_c$. 

This visualization is used only as empirical support for the covariance-sparsification behavior; it is not a proof of universal diagonalization or a direct prediction of bitrate reduction. The quantitative evidence for coding gain is provided by the controlled ablations in Table~\ref{ablation_different_modules}.

We further visualize the ChWP entropy layout because its role is different from the $\mathrm{WT}_c/\mathrm{IWT}_c$ wrapper in ChWDTB. The previous visualization focuses on the feature coordinates seen by ChWDTA, whereas ChWP directly determines the order and grouping of the latent variables coded by ChARM. For this reason, the ChWP visualization is constructed at the entropy-slice level rather than at the individual-channel level. We use Kodak image~08 as an illustrative image-level case study because it clearly exhibits the slice-structure induced by the two-level ChWP transform. This single-image case is used only for interpretation; the coding gain of ChWP is still evaluated by the controlled Kodak ablation in Table~\ref{ablation_different_modules} and the entropy-grouping comparison in Table~\ref{tab:ablate_depth}.

\begin{figure*}[t]
  \centering
  \includegraphics[width=\textwidth]{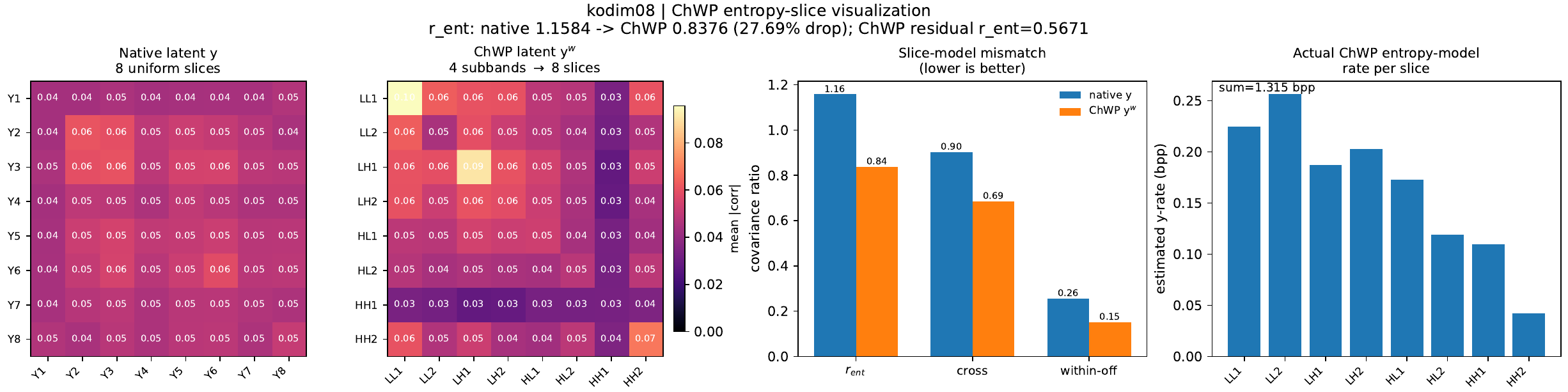}
  \caption{Entropy-slice visualization of the two-level ChWP transform on Kodim~08 from Kodak dataset. The first panel shows the slice-level mean absolute correlation when the native latent $\mathbf{y}$ is uniformly divided into eight channel slices. The second panel shows the corresponding result after ChWP, where the eight slices are ordered as $\mathrm{LL}_1$, $\mathrm{LL}_2$, $\mathrm{LH}_1$, $\mathrm{LH}_2$, $\mathrm{HL}_1$, $\mathrm{HL}_2$, $\mathrm{HH}_1$, and $\mathrm{HH}_2$. The third panel compares the covariance-based entropy sparsity ratio $r_{\mathrm{ent}}$ in Eq.~\eqref{eq:entropy_sparsity} and its two components: the cross-slice covariance term and the within-slice off-diagonal term. The fourth panel reports the estimated entropy-model rate of each ChWP slice. For this image, $r_{\mathrm{ent}}$ decreases from $1.1584$ to $0.8376$, corresponding to a $27.69\%$ reduction. The cross-slice component decreases from $0.90$ to $0.69$, and the within-slice off-diagonal component decreases from $0.26$ to $0.15$. The ChWP residual ratio after entropy prediction is $0.5671$, and the total estimated $y$-rate is $1.315$~bpp.}
  \label{fig:chwp_entropy_vis}
\end{figure*}

Fig.~\ref{fig:chwp_entropy_vis} supports the entropy-model interpretation in Sec.~\ref{sec:entropy}. Compared with directly splitting the native latent $\mathbf{y}$ into eight uniform slices, the ChWP-domain representation $\mathbf{y}^{w}$ produces a slice layout with a lower covariance mismatch ratio. The reduction of $r_{\mathrm{ent}}$ from $1.1584$ to $0.8376$ indicates that the ChWP ordering leaves less covariance structure outside the modeling assumptions of the slice-wise diagonal likelihood. Importantly, both terms in Eq.~\eqref{eq:entropy_sparsity} are reduced: the cross-slice term drops from $0.90$ to $0.69$, showing weaker dependence between sequentially coded slices, while the within-slice off-diagonal term drops from $0.26$ to $0.15$, showing that each slice becomes closer to the diagonal covariance structure assumed by the Gaussian likelihood. The heatmaps also show that ChWP does not simply suppress every correlation entry uniformly. Instead, it reorganizes the latent into interpretable subband-like groups: the LL/LH slices retain more energy and rate, while the HH-related slices have lower mean correlation and lower coding rate. This behavior is consistent with the design goal of ChWP: it provides a balanced eight-slice ordering that is easier for ChARM to model sequentially, rather than serving as an exact diagonalization transform. Therefore, Fig.~\ref{fig:chwp_entropy_vis} should be interpreted as qualitative support for the slice-alignment argument, while the BD-rate contribution of ChWP is quantified by the ablation studies.

\subsection{Ablation Studies}

\subsubsection{Effect of Proposed Modules}

Unless otherwise stated, all ablations follow the same training and evaluation protocol as the full model and report Kodak BD-rate with VVC as the anchor.

To assess the contribution of the wavelet-domain attention path, we first remove $\mathrm{WT}_c$ and $\mathrm{IWT}_c$ in ChWDTB while keeping all other components unchanged. We denote this variant as \textbf{w/o TB Lift}. We then examine the contribution of the channel-wise wavelet packet \textbf{ChWP} by removing this module as well. In this case, the latent representation $\mathbf{y}$ is uniformly partitioned into eight slices directly and the remaining design is left unchanged. We denote this variant as \textbf{w/o TB Lift, ChWP}. The results are summarized in Table~\ref{ablation_different_modules}.

\begin{table}[t]
\centering
\caption{Ablation on the wavelet-domain attention and ChWP entropy coding under the default 8-slice setting.}
\label{ablation_different_modules}
\begin{tabular}{lccc}
\hline
Method & \textbf{Latency} & \textbf{Params}  & \textbf{BD-rate} \\
\hline
\textbf{Ours (8-slice)}  & 225 ms & 189 M & $-17.82\%$ \\
\textbf{w/o TB Lift} & 221 ms & 189 M & $-16.76\%$ \\
\textbf{w/o TB Lift, ChWP} & 218 ms & 189 M  & $-15.50\%$ \\
\hline
\end{tabular}
\end{table}

Compared with the full 8-slice method, removing $\mathrm{WT}_c/\mathrm{IWT}_c$ from the attention path degrades the Kodak BD-rate by 1.06 percentage points with almost unchanged latency and no change in parameters. Further removing ChWP degrades it by another 1.26 percentage points with only a small latency reduction and the same parameter count. These results show that both the wavelet-domain attention and the ChWP entropy path contribute meaningful coding gains at essentially constant model size.

\subsubsection{Effect of Entropy Grouping in the Entropy Model}

We compare the lightweight 4-slice entropy layout and the default 8-slice entropy layout while keeping the two-level ChWP transform unchanged. The 4-slice model directly uses the four ChWP subbands as four sequential entropy slices. The 8-slice model further splits each subband into two channel slices, resulting in eight sequential slices and a stronger conditional entropy model.

Table~\ref{tab:ablate_depth} reports the results on Kodak. The 8-slice model improves the Kodak BD-rate by 0.58 percentage points over the 4-slice model, at the cost of higher entropy-model complexity. Specifically, the latency increases from 187~ms to 225~ms, kMACs per pixel increase by 31.8\%, and parameters increase by 67.3\%. We therefore use the 8-slice model as the default setting for the main R--D comparison, because it gives the best compression performance. The 4-slice model is retained as a hardware-friendly variant for applications where memory and computation are more constrained.

\begin{table}[t]
\centering
\caption{Ablation on entropy grouping in the ChWP path on Kodak.}
\label{tab:ablate_depth}
\begin{tabular}{lcccc}
\hline
Setting & Latency & kMACs/pixel & Params & BD-rate \\
\hline
4-slice, $S=4$  & 187 ms & 914  & 113 M & $-17.24\%$ \\
8-slice, $S=8$  & 225 ms & 1205 & 189 M & $\mathbf{-17.82\%}$ \\
\hline
\end{tabular}
\end{table}

\subsubsection{Effect of $\mathrm{WT}_c$ Levels in ChWDTB}

We vary the number of $\mathrm{WT}_c$ levels in ChWDTB while keeping all other components fixed. Table~\ref{tab:levels_in_chwdtb} reports results on Kodak.

\begin{table}[t]
\centering
\caption{Ablation on the $\mathrm{WT}_c$ levels in ChWDTB under the default 8-slice setting.}
\label{tab:levels_in_chwdtb}
\begin{tabular}{ccccc}
\hline
Levels & Latency & kMACs/pixel & Params & BD-rate \\
\hline
1  & 225 ms & 1205  & 189 M & $-17.82\%$ \\
2  & 228 ms & 1205  & 189 M & $-17.83\%$ \\
\hline
\end{tabular}
\end{table}

On Kodak, two $\mathrm{WT}_c$ levels give essentially the same BD-rate as one level. The parameters and kMACs/pixel are unchanged, but the latency increases from 225~ms to 228~ms. We therefore choose one level for a better accuracy--latency trade-off. This is consistent with the role of $\mathrm{WT}_c$ in ChWDTB as an invertible reparameterization before attention: one level already forms useful wavelet-induced channel branches, and extra levels bring only negligible R--D gains.

\subsubsection{Effects of Different Wavelets}

We compare Haar, CDF~9/7 (bior4.4), and learned wavelets in ChWDTB and ChWP. In this ablation, one module is varied at a time while the other modules use the default learned bior4.4 setting.

\begin{table}[t]
\centering
\caption{Effects of wavelets in ChWDTB and ChWP using the Kodak set.}
\label{tab:ablate_wavelet_chwdtb}
\small
\setlength{\tabcolsep}{6pt}
\begin{tabular}{c c c c}
\toprule
Module & Wavelet & BD-rate (\%) & Latency (ms) \\
\midrule
ChWP   & Haar (fixed)        & $-16.71$ & 218 \\
ChWP   & bior4.4 (fixed)     & $-16.99$ & 223 \\
ChWP   & bior4.4 (learned)   & \textbf{$-17.82$} & 225 \\
\midrule
ChWDTB & Haar (fixed)        & $-17.71$ & 223 \\
ChWDTB & bior4.4 (fixed)     & $-17.81$ & 225 \\
ChWDTB & bior4.4 (learned)   & $-17.82$ & 227 \\
\bottomrule
\end{tabular}
\end{table}

Small latency differences of 1--2~ms for nearby variants in this ablation are within the measurement variation of repeated GPU runs; the BD-rate trend is therefore the main comparison target.

Table~\ref{tab:ablate_wavelet_chwdtb} leads to three observations. First, CDF~9/7 is better than Haar in both modules, suggesting that the smoother biorthogonal lifting template is a better initialization for the learned channel reparameterization. Second, the learned lifting wavelet gives the best BD-rate in both places. Third, the gain is larger for ChWP than for ChWDTB. This asymmetry is consistent with the restricted-model view: in ChWP the wavelet directly determines the subband layout used by the hyperprior and the slice-based discretized Gaussian model, so learning can better align the slices to the empirical latent distribution. In ChWDTB, the wavelet mainly reparameterizes features around attention, and a fixed bior4.4 already captures most of the benefit. Therefore, module-specific learned wavelets are justified when they improve validation BD-rate, while a mixed design with a fixed backbone wavelet and a learned entropy-path wavelet remains a reasonable accuracy--interpretability trade-off.

\subsubsection{Attention Networks in the Entropy Model}

We compare the proposed ChWDTB-based \textit{AttentionNet} with a residual block-based \textit{AttentionNet} following~\cite{Fu_EECV} under identical 8-slice settings on Kodak. As shown in Table~\ref{tab:different_net}, the ChWDTB-based design achieves an additional $0.56$ percentage-point BD-rate reduction at comparable complexity, indicating that wavelet-domain attention benefits entropy parameter estimation with minimal overhead.

\begin{table}[t]
\centering
\caption{Comparison of attention networks in the entropy model on Kodak.}
\label{tab:different_net}
\begin{tabular}{lcccc}
\hline
Method & Latency  & kMACs/pixel & Params & BD-rate \\
\hline
\textbf{Ours (8-slice)}         & 225 ms & 1205 & 189 M & $-17.82\%$ \\
\cite{Fu_EECV} & 222 ms & 1203  & 186 M & $-17.26\%$ \\
\hline
\end{tabular}
\end{table}

\section{Conclusion}

In this paper, we develop an improved hybrid CNN--transformer learned image compression scheme based on channel-wise wavelet-domain sparsification. Different from existing methods that apply spatial attention directly in the native feature coordinates, the proposed ChWDTB computes spatial attention through ChWDTA from wavelet-transformed channel coordinates and then maps the result back with $\mathrm{IWT}_c$. The channel-wise transform is reversible, so its role is not to discard information; rather, it sparsifies the channel covariance seen by finite-capacity attention and entropy models. We further perform entropy coding after a learned channel-wise wavelet packet (ChWP) decomposition, which produces balanced subbands and slices that better match the slice-wise diagonal Gaussian likelihood used by ChARM. The same ChWDTB is also integrated into the entropy path and the hyperprior.

Experimental results on Kodak, CLIC Professional Validation, and Tecnick show that the default 8-slice model obtains BD-rate values of $-17.82\%$, $-19.15\%$, and $-22.56\%$, respectively, which are the most favorable BD-rate results among the compared codecs in our evaluation. Relative to DCAE, the BD-rate margins are $0.84$, $2.17$, and $2.43$ percentage points on the three datasets, with larger gains on the two higher-resolution benchmarks. The lightweight 4-slice variant obtains BD-rate values of $-17.24\%$, $-18.42\%$, and $-21.71\%$ while using fewer parameters and fewer kMACs/pixel than DCAE, with a comparable total latency on Kodak. These results demonstrate that structured channel-wise wavelet-domain sparsification can improve compression performance while providing a controllable accuracy--complexity trade-off.

Beyond the numerical gains, the significance of this study is that it shows how classical transform-coding principles can be reintroduced into modern learned codecs as reversible channel-wise reparameterizations rather than as hand-crafted spatial-frequency partitions. This perspective provides a practical way to improve the compatibility between transformer backbones, slice-wise entropy models, and GPU-friendly parallel decoding. Future work will investigate content-adaptive or rate-adaptive channel-wise wavelet bases, extend the same reparameterization principle to perceptual, variable-rate, and video compression settings, and develop a tighter theoretical connection between covariance sparsification, entropy-model mismatch, and realized bit-rate savings.



\end{document}